\documentclass[aps,prx,twocolumn,superscriptaddress,
 amsmath,amssymb,amsfonts,floatfix]{revtex4-2}
\usepackage{microtype}
\usepackage{graphicx}
\usepackage{bm}
\usepackage[dvipsnames]{xcolor}
\usepackage{hyperref}
\usepackage{tikz}
\usetikzlibrary{decorations.pathmorphing} 
\usetikzlibrary{decorations.markings} 
\tikzset{
  midarrowboth/.style={
    postaction={decorate},
    decoration={
      markings,
      mark=at position 0.3 with {\arrow{>}},
      mark=at position 0.7 with {\arrow{<}}
    }
  }
}
\tikzset{
  midarrow/.style={
    postaction={decorate},
    decoration={markings, mark=at position 0.5 with {\arrow{>}}}
  }
}
\hypersetup{
    colorlinks=true,
    citecolor=Green,
    linkcolor=Red,
    urlcolor=NavyBlue
}
\usepackage[capitalise]{cleveref}

\usepackage{soul}
\usepackage[normalem]{ulem}
\usepackage{mathtools}
\usepackage{siunitx}

\definecolor{hotpink}{RGB}{255, 105, 180}   

\newcommand{\LL}{\mathcal{\hat L}}

\DeclarePairedDelimiter\bra{\langle}{\rvert}
\DeclarePairedDelimiter\ket{\lvert}{\rangle}
\begin{document}
\title{What Hall Measures in Strongly Correlated Metals}
\author{Emily~Z.~Zhang}
\affiliation{Stanford Institute for Materials and Energy Sciences,
SLAC National Accelerator Laboratory, 2575 Sand Hill Road, Menlo Park, CA 94025, USA}
\affiliation{Geballe Laboratory for Advanced Materials, Stanford University, California 94305, USA}

\author{Thomas P. Devereaux}
\email{tpd@stanford.edu}
\affiliation{Stanford Institute for Materials and Energy Sciences,
SLAC National Accelerator Laboratory, 2575 Sand Hill Road, Menlo Park, CA 94025, USA}
\affiliation{
Department of Materials Science and Engineering, Stanford University, Stanford, CA 94305, USA}
\affiliation{Geballe Laboratory for Advanced Materials, Stanford University, California 94305, USA}
\begin{abstract}
The Hall effect provides a sensitive probe of electronic correlations in strongly interacting metals, yet its microscopic interpretation beyond quasiparticle transport remains poorly understood. Here, we develop a Liouvillian formulation of the Hall conductivity that organizes the response into different classes of flux-sensitive many-body interference operators. Using a Schrieffer–Wolff transformation of the Hubbard model, we identify how projected charge motion and virtual hopping processes contribute to oriented trajectories enclosing magnetic flux. A high-temperature expansion reveals a hierarchy of inverse-temperature contributions associated with these interference processes. This framework provides a possible microscopic interpretation of the integer power laws previously observed in determinant quantum Monte Carlo calculations, which emerge at temperatures below the regime where the expansion is formally controlled. Our results highlight the sensitivity of Hall transport to coherent many-body dynamics and establish a connection between lattice geometry, electronic correlations, and transverse charge response.
\end{abstract}

\date{\today}

\maketitle
\section{Introduction}
The Hall effect occupies a unique position among probes of condensed matter. Unlike the longitudinal conductivity, which measures the ability of charge to propagate through a material, the Hall response requires carriers to acquire a transverse deflection generated by time reversal symmetry breaking. In the case of conventional metals, the Hall coefficient is determined primarily by the Fermi surface geometry~\cite{ong_geometric_1991}, while the longitudinal conductivity depends on the carrier density and scattering rate. Together, these quantities provide a direct experimental window into quasiparticle dynamics. 

Strongly correlated metals present a fundamentally different situation. In cuprates~\cite{gurvitch_resistivity_1987,martin_temperature_1988,takagi_systematic_1992}, heavy fermion compounds~\cite{custers_break-up_2003,custers_evidence_2010,nakajima_non-fermi_2007,paschen_hall-effect_2004,tomita_strange_2015,taupin_are_2022}, twisted moiré materials~\cite{cao_strange_2020,lyu_strange_2021,xia_bandwidth-tuned_2026}, nickelates~\cite{li_superconducting_2020}, and numerous other systems~\cite{phillips_stranger_2022-1}, transport occurs in regimes where quasiparticles become poorly defined, resistivity exceeds the Mott-Ioffe-Regel limit~\cite{gurvitch_resistivity_1987,emery_superconductivity_1995-1,gunnarsson_colloquium_2003-1,hussey__universality_2004-1}, and conventional Boltzmann theory loses its microscopic foundation. Although Hall effect measurements are commonly performed in these materials, their interpretations remain ambiguous. Hall coefficients frequently change sign~\cite{tsukada_negative_2006,ando_evolution_2004}, vary strongly with temperature~\cite{chien_effect_1991,hwang_scaling_1994,ono_strong_2007}, and exhibit material-dependent behavior that cannot be understood solely from carrier density or Fermi-surface topology~\cite{ando_nonuniversal_1999,carrington_temperature_1992,putzke_reduced_2021,hayes_superconductivity_2021}. At the same time, the longitudinal resistivity often displays remarkably universal behavior, most notably the ubiquitous linear-in-temperature resistivity associated with strange metallic transport~\cite{bruin_similarity_2013,legros_universal_2019,hartnoll_scaling_2015-1}, which is also found in unbiased calculations and quantum simulations of the Hubbard model~\cite{huang_strange_2019-3,brown_bad_2019}.

This contrast illustrates the lack of understanding of the Hall response once the quasiparticle picture breaks down. Existing theories typically attempt to generalize semiclassical transport by introducing modified scattering rates~\cite{stojkovic-acute_theory_1997,abrahams_hall_2003}, current vertices~\cite{kontani_anomalous_2008}, or phenomenological lifetimes~\cite{anderson_hall_1991-1,coleman_how_1996-1}. Approaches that avoid quasiparticles altogether have related the Hall response to equilibrium or short-time quantities, including the high-frequency Hall constant~\cite{shastry_faraday_1993}, exact equilibrium formulas for the Hall number~\cite{auerbach_hall_2018}, and memory-matrix descriptions of magnetotransport~\cite{lucas_memory_2015}. Unbiased numerical calculations of the Hall coefficient in the Hubbard model have also become available~\cite{wang_dc_2020,wang_numerical_2021-1}. While these approaches can reproduce aspects of experiments, it remains an open question of whether the Hall conductivity measures the same microscopic physics as its longitudinal counterpart, or whether it probes an entirely different aspect of correlated electronic motion. 

The distinction between longitudinal and transverse transport can be understood by examining the nature of the collective charge carriers. The longitudinal conductivity is nonzero when charge carriers propagate between neighboring sites. By contrast, a Hall current requires an electronic motion that is sensitive to magnetic flux, generated intrinsically or extrinsically. This dependence on flux demands interference between alternative hopping trajectories that enclose a finite area~\cite{aharonov_significance_1959}. The Hall conductivity therefore probes many-body interference processes beyond those governing ordinary charge transport. Understanding precisely which correlations these are and how they emerge from a microscopic Hamiltonian has remained an outstanding theoretical problem.

Recently, we demonstrated numerically that the Hall conductivity provides a sensitive probe of the crossover from semi-classical to quantum-coherent transport in the doped Hubbard model, while the longitudinal resistivity remains largely insensitive to this crossover~\cite{zhang_hall_2026-1}. Our previous work showed that the Hall response is governed by a sequence of integer power laws that emerge as quantum coherence develops. In particular, the Hall response exhibits distinct integer power laws whose onset tracks the emergence of a coherence scale extracted independently from double occupancy. These observations suggested that Hall transport is governed not only by charge motion, but by coherent flux-enclosing hopping processes that develop as many-body coherence emerges. The numerical results, however, pose a fundamental question: Why is the Hall conductivity sensitive to quantum coherence, in contrast to the longitudinal channel? 

In this work, we develop a microscopic theory of Hall transport directly from the Kubo formalism~\cite{kubo_statistical_1957,thouless_quantized_1982} without invoking quasiparticles or Boltzmann transport. We derive a formal inverse-temperature expansion of the Hall conductivity in terms of equilibrium correlation functions,
\[
\sigma_{xy}
=
\frac{C_1}{T}
+
\frac{C_2}{T^2}
+
\frac{C_3}{T^3}
+\cdots,
\]
where each coefficient $C_i$ is the thermal expectation value of a progressively more complex interference operator. In the strong-coupling limit~\cite{macdonald_tu_1988}, these operators reduce to coherent flux-enclosing loops. Unlike conventional high-temperature expansions~\cite{shastry_faraday_1993,perepelitsky_transport_2016}, these coefficients are expressed as expectation values of operators that count coherent flux-enclosing hopping processes. The expansion therefore identifies the Hall effect not as a measure of carrier density or scattering lifetimes, but as a hierarchy of increasingly complex many-body loop correlations. We show how lattice geometry and particle-hole symmetry constrain the allowed flux-enclosing interference processes. In particular, next-nearest-neighbor hopping introduces triangular paths that are absent in the nearest-neighbor square-lattice model, thereby modifying the microscopic contributions to Hall transport. More generally, we illuminate how the integer power law exponent is dependent on the class of coherent hopping loop dominating transport.

Our results provide an alternative framework for interpreting Hall transport in correlated matter. Whereas the longitudinal conductivity primarily measures the existence of mobile charge carriers, the Hall conductivity measures the emergence of coherent electronic motion around closed paths that accumulate magnetic phase. In this sense, the Hall effect provides a direct experimental probe of many-body quantum coherence itself. Instead of merely serving as a geometric property of a Fermi surface, the Hall response becomes a quantitative measure of the correlated electronic loops that develop as an incoherent metal evolves toward a quantum-coherent state.

\section{Liouvillian formulation of Hall transport}
\label{sec:Louie}
The conductivity tensor is an equilibrium property of the many-body system and is therefore determined by linear response theory. Throughout this work, we make no assumptions regarding quasiparticles, semiclassical transport, or the existence of well-defined electronic excitations. Instead, our objective is to determine directly the microscopic operator measured by a Hall experiment.

For an arbitrary interacting Hamiltonian $H$, the optical conductivity is
obtained from the Kubo formula
\begin{equation}
\sigma_{\alpha\beta}(z)
=
\frac{i}{Vz}
\left[
\langle \hat \tau_{\alpha\beta}\rangle
+
\chi_{\alpha\beta}^{R}(z)
\right],
\label{eq:kubo}
\end{equation}
where
\begin{equation}
\chi_{\alpha\beta}^{R}(t)
=
-\frac{i}{\hbar}
\Theta(t)
\left<
[\hat J_\alpha(t),\hat J_\beta(0)]
\right>
\label{eq:retarded}
\end{equation}
is the retarded current-current correlation function and
$\hat \tau_{\alpha\beta}$ is the diamagnetic stress tensor operator. 

To probe the dynamics of the current operators, we introduce the Liouvillian superoperator
\begin{equation}
\mathcal{\hat L}\hat O
=
[\hat H,\hat O],
\label{eq:Liouvillian}
\end{equation}
which generates dynamics directly in operator space. Unlike the Hamiltonian,
which propagates many-body wavefunctions, the Liouvillian propagates operators.
The Hall conductivity may therefore be viewed as measuring the evolution of the
current operator within this operator space.

Using the identity
\begin{equation}
\hat J_\alpha(t)
=
e^{i\mathcal{\hat L}t/\hbar}
\hat J_\alpha,
\end{equation}
the retarded response function may be written exactly as
\begin{equation}
\chi_{\alpha\beta}^{R}(z)
=
\frac{1}{V}
\left<
\hat J_\alpha
\frac{1-e^{-\beta\mathcal{\hat L}}}
{\hbar z-\mathcal{\hat L}}
\hat J_\beta
\right>_\beta,
\label{eq:chi_liouvillian}
\end{equation}
where $\langle \cdot \rangle_\beta=\frac{1}{Z}\text{Tr}[e^{-\beta\hat{H}} \cdot ]$ is the thermal expectation value with respect to the equilibrium density matrix. 

We are interested in the DC conductivity $\sigma_H=\frac{1}{2}(\sigma_{xy}-\sigma_{yx}$ as $z\to 0$. The resolvent from \eqref{eq:chi_liouvillian} for small $z$ can be expanded as 
\begin{align}
    \frac{1}{\hbar z-\LL} =-\sum_{k=0}^\infty(\hbar z)^k\LL^{-k-1}.  
\end{align}
Inserting this series back into \eqref{eq:chi_liouvillian} produces a Laurent series in $z$,
\begin{align}
    \sigma_H=\frac{1}{Vz}\sum_{k=0}^\infty (\hbar z)^k\ \mathrm{Im}\langle\hat J_x (1-e^{-\beta\LL}) \LL^{-k-1}\hat J_y\rangle_\beta\label{eq:small_z_sigma}.
\end{align}
One can show that the apparent singularity $1/z$ at $k=0$ vanishes (see Appendix \ref{appendixc}). The $k=1$ term then yields the dc Hall conductivity
\begin{align}
    \mathrm{Re}\ \sigma_H = \frac{\hbar}{V}\mathrm{Im}\left< \hat J_x \frac{1-e^{-\beta\LL}}{\LL^2} \hat J_y\right>_\beta,\label{eq:dc_exact}
\end{align}
which is equivalent to the usual Lehmann spectral representation expressed in terms of Liouvillian operators.

\subsection{High temperature hierarchy of interference operators}
Up until this section, the Liouvillian formulation of the dc Hall conductivity is exact and general on the operator level. We now examine how the Hall conductivity organizes itself at high temperature. We begin by expanding the thermal factor in \eqref{eq:dc_exact}, 
\begin{align}
    \frac{1-e^{-\beta\LL}}{\LL^2}&=\sum_{p=1}^\infty \frac{(-1)^{p+1}\beta^p}{p!}\LL^{p-2}\\
    &=\beta \LL^{-1} - \frac{\beta^2}{2}+\frac{\beta^3}{6}\LL-\cdots .
\end{align}
Substituting this expansion back into \eqref{eq:dc_exact} gives 
\begin{equation}
    \mathrm{Re}\ \sigma_H = \frac{\hbar}{V} \sum_{p=1}^\infty \frac{(-1)^{p+1}\beta^p}{p!} \mathrm{Im}\langle\hat J_x \LL^{p-2}\hat J_y  \rangle_\beta. 
\end{equation}
This expression reveals that successive terms correspond to increasingly complicated nested commutators 
\begin{align}
\mathcal{\hat L}\hat J_y
&=
[\hat H,\hat J_y],
\\
\mathcal{\hat L}^2\hat J_y
&=
[\hat H,[\hat H,\hat J_y]],
\\
\mathcal{\hat L}^3\hat J_y
&=
[\hat H,[\hat H,[\hat H,\hat J_y]]],
\end{align}
and so forth. In other words, each successive application of the Liouvillian increases the spatial complexity of the operator generated from the current. The Hall conductivity therefore organizes itself according to a hierarchy of increasingly complex many-body operators. 

The thermal expectation value has its own expansion. Taking the $\beta$ dependence of the partition function into account, 

\begin{align*}
\langle \mathcal{\hat A}\rangle_\beta
=&
\langle \mathcal {\hat A}\rangle_\infty \\
&-
\beta
\left[
\langle \hat H \mathcal{\hat A}\rangle_\infty
-
\langle \hat H\rangle_\infty
\langle \mathcal{\hat A}\rangle_\infty
\right] \\
&+\frac{\beta^2}{2}\Biggl[\langle \hat H^2\mathcal{\hat A}\rangle_\infty -2\langle \hat H\rangle_\infty\langle\hat H\mathcal{\hat A}\rangle_\infty \\
&-\langle\hat H^2\rangle_\infty\langle \mathcal{\hat A}\rangle_\infty+2\langle\hat H\rangle^2_\infty\langle\mathcal{\hat A}\rangle_\infty
\Biggr]+\cdots,
\end{align*}

where
\begin{equation*}
\langle \hat A\rangle_\infty
=
\frac{1}{\mathcal N}
\mathrm{Tr}\,\hat A
\end{equation*}
denotes the infinite-temperature average over the relevant Hilbert space having dimension ${\mathcal N}$. Combining this expansion yields a final expression revealing an operator hierarchy of the form 

\begin{equation}
\sigma_{H}
=
\frac{C_1}{T}
+
\frac{C_2}{T^2}
+
\frac{C_3}{T^3}
+\cdots .
\label{eq:master}
\end{equation}
Equation \eqref{eq:master} represents the central organizing principle of this work. The coefficients $C_n$ are equilibrium expectation values of progressively more
complex interference operators,
\begin{widetext}
\begin{align}
C_1&=
\frac{\hbar}{V}\mathrm{Im} 
\langle
\hat J_x \mathcal{\hat L}^{-1}\hat J_y\rangle_\infty,\label{eq:C1}\\
C_2&=-\frac{\hbar}{2V}\mathrm{Im}\langle\hat J_x \hat J_y\rangle_\infty-\frac{\hbar}{V}\mathrm{Im} 
\langle \hat H
\hat J_x \mathcal{\hat L}^{-1}\hat J_y\rangle_\infty+\frac{\hbar}{V}\langle \hat H\rangle_\infty\mathrm{Im} 
\langle
\hat J_x \mathcal{\hat L}^{-1}\hat J_y\rangle_\infty,\label{eq:C2}\\
C_3&=\frac{\hbar}{6V}\mathrm{Im}\langle
    \hat J_x \mathcal{\hat L}\hat J_y\rangle_\infty
    +\frac{\hbar}{2V}\mathrm{Im}\langle \hat H\hat J_x \hat J_y\rangle_\infty
-\frac{\hbar}{2V}\langle\hat H\rangle_\infty\mathrm{Im}\langle\hat J_x \hat J_y\rangle_\infty
+\frac{\hbar}{2V}\mathrm{Im} 
\langle \hat H^2
\hat J_x \mathcal{\hat L}^{-1}\hat J_y\rangle_\infty\nonumber\\ &-\frac{\hbar}{V}\langle\hat H\rangle_\infty\mathrm{Im} 
\langle \hat H
\hat J_x \mathcal{\hat L}^{-1}\hat J_y\rangle_\infty
-\frac{\hbar}{2V}\langle\hat H^2\rangle_\infty
\mathrm{Im} 
\langle
\hat J_x \mathcal{\hat L}^{-1}\hat J_y\rangle_\infty
+\frac{\hbar}{V}\langle\hat H\rangle^2_\infty
\mathrm{Im} 
\langle
\hat J_x \mathcal{\hat L}^{-1}\hat J_y\rangle_\infty.\label{eq:C3}
\end{align}

\end{widetext}

Computing these coefficients thus amounts to evaluating terms of the form 
\begin{equation}
    \frac{1}{V}\text{Im}\langle \hat J_x \LL^k \hat J_y\rangle_\infty =\frac{1}{\mathcal N}\sum_{nm,n\ne m}(E_n-E_m)^k\ \text{Im}[\hat j_{nm}^x\hat j^y_{mn}], \label{eq:jj_matrix}
\end{equation}
where $n,m$ label the eigenstates, $E_{n}$ and $E_m$ are the eigenenergies, and the matrix elements for the current operator are given by $\hat j_{nm}=\bra{n}{\hat j}\ket{m}$. 

These terms are non-zero only if the infinite-temperature trace contains a closed, oriented, gauge-invariant loop. Physically, these coefficients in Eq.~(\ref{eq:master}) measure distinct interference operators, whose microscopic structures are discussed in Section~\ref{sec:largeU}. 

This hierarchy of coefficients illuminates why the Hall conductivity exhibits extended integer power laws over broad temperature ranges. Whenever one interference
operator dominates over the others,
\[
\frac{|C_n|}{T^n}
\gg
\frac{|C_m|}{T^m},
\qquad
m\neq n,
\]
the Hall conductivity reduces to
\[
\sigma_{H}
\propto
T^{-n}.
\]

If one term in the inverse-temperature expansion dominates over a given temperature interval within its regime of validity, the Hall conductivity exhibits the corresponding integer power law. Whether this hierarchy also governs the lower-temperature power laws observed numerically remains an important question, which we revisit in Sec. \ref{sec:discussion}.

For completeness, Appendix~\ref{appendixd} shows that the interference picture
developed here is already implicit in the Kubo formula, and not dependent upon
the Liouvillian reorganization introduced here.

\section{Strong-coupling representation of the Hall operator}
\label{sec:largeU}

The Hall conductivity is determined by the interference
between competing many-body hopping trajectories irrespective of whether the
electrons are weakly or strongly correlated. Strong correlations, however,
drastically modify the trajectories that remain dynamically accessible. To understand the physical content of these interference operators, it is useful to specialize to the strongly interacting Hubbard model, where their structure becomes especially transparent. 

To illustrate the strong-coupling effects, we examine the $t$-$t'$-$U$ Hubbard-Hofstader model on the square lattice given by
\begin{equation}
    \hat H=\sum_{ij\sigma}t_{ij}\hat c^\dagger_{ij}\hat c_{j\sigma}+U\sum_i\hat n_{i\uparrow}\hat n_{i\downarrow}, 
\end{equation}
where $t_{ij}$ is either nearest neighbour $t$ or next-nearest-neighbour $t'$ and encodes a Peierls phase $e^{iA_{ij}}$, and $U$ is the Coulomb repulsion energy. In the limit where $U \gg t,t'$, the electronic motion separates into low-energy states containing singly occupied sites and high-energy states involving doubly occupied configurations. Since the Coulomb interaction penalizes double occupancy by an energy $U$, charge transport proceeds primarily through virtual excursions into the upper Hubbard band. The Hall conductivity therefore probes the interference between these virtual hopping processes.

Because $U$ is the dominant energy scale, we project the Hamiltonian and the current operators into the manifold containing no doubly occupied sites using a Schrieffer-Wolff transformation,
\begin{align}
    \hat H_{\rm eff} & = e^{\hat S}\hat H e^{-\hat S},\\
    \hat{j}^{\text{eff}}_\alpha &= \hat P e^{\hat S} \hat{j}_\alpha e^{-\hat S}\hat P
\end{align}
where the anti-Hermitian generator $\hat S$ removes matrix elements connecting different Hubbard sectors order-by-order in $t/U$, and $\hat P$ projects onto the low-energy manifold. We can then compute the matrix elements $\text{Im}[\hat j_{nm}^x\hat j^y_{mn}]$ in Eq. \eqref{eq:jj_matrix} in this projected subspace.

This projection is achieved by first re-expressing the full Hamiltonian as $\hat H=\hat H_d+\hat H_{KE}$, where $\hat H_d=U\hat D=U\sum_i\hat n_{i,\uparrow}\hat n_{i,\downarrow}$ is diagonal in doublon number, and the kinetic energy $\hat H_{KE}=\hat T_{-1}+\hat T_0+\hat T_1$ has three parts that conserves the number of doublon-hole pairs, or increases/decreases the doublon-hole pairs by one. $\hat S$ can then be expressed as $\hat S=\frac{1}{U}[\hat T_{+1}-\hat T_{-1}]$, and is perturbatively applied to yield both $\hat H_\text{eff}$ and the projected current operator $\hat{j}^{\text{eff}}_\alpha$.
Performing this transformation yields the expansions
\begin{align*}
\hat H_\text{eff}&= U\hat D + \hat T_0+\frac{1}{U}[\hat T_{+1},\hat T_{-1}] +\dots\\
\hat{j}^{\text{eff}}_\alpha &=\hat P\hat j_\alpha\hat P + \hat P [\hat S, \hat j_\alpha]\hat P+\frac{1}{2}\hat P[\hat S,[\hat S,\hat j_\alpha]]\hat P+\dots.
\end{align*}
Then,
\begin{align}
\hat j^\text{eff}_\alpha = \hat j_0+\frac{1}{U}\hat P[\hat T_{+1}-\hat T_{-1},\hat j_\alpha]\hat P +\dots
\label{eq:v_pert}
\end{align}
where $\hat j_0$ connects states having the same number of doublon-hole pairs. The effective current operator acquires higher-order terms,
\begin{equation*}
\hat j^\text{eff}_\alpha
=
\hat j_\alpha^{(1)}
+
\hat j_\alpha^{(2)}
+
\cdots ,
\end{equation*}
with
\begin{equation*}
\hat j_\alpha^{(1)} \sim t_{ij},
\qquad
\hat j_\alpha^{(2)} \sim \frac{t_{ij}^2}{U}.
\end{equation*}

This operator contains both direct hopping processes within the projected low-energy manifold and higher-order contributions arising from virtual excursions into states with double occupancy. At finite doping, the leading projected current permits hole motion without creating doublons and can therefore contribute to flux-enclosing interference processes already at zeroth order in \(1/U\). Such contributions are distinct from the virtual doublon-mediated processes generated by the Schrieffer–Wolff transformation. The latter provide additional flux-sensitive terms in the effective current, with amplitudes of order \(t^2/U\), and modify the Hall response through interference with the leading projected hopping current. Consequently, the triangular and square loops identified below represent particular strong-coupling contributions to the Hall response, rather than necessarily its leading contributions at finite doping.

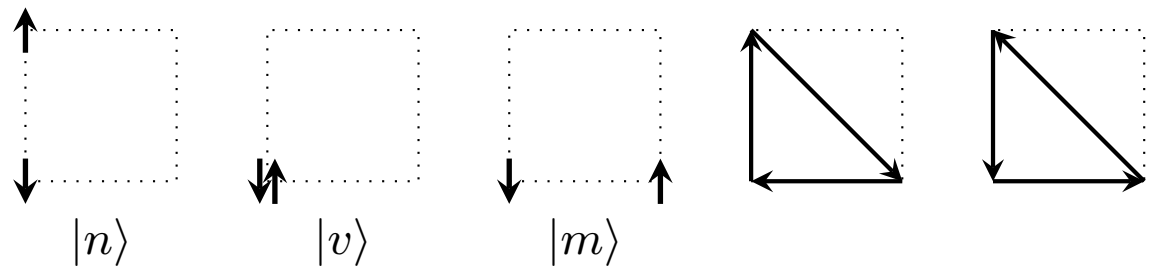
\begin{figure}
    \centering
\begin{tikzpicture}[scale=1.0, >=stealth]

\newcommand{\drawstate}[3]{
  \begin{scope}[shift={(#1,0)}]

    \draw[dotted] (0,0) -- (1,0) -- (1,1) -- (0,1) -- cycle;

    #3

    \node at (0.5,-0.4) {#2};

  \end{scope}
}

\drawstate{0}{$|n\rangle$}{
  \draw[->, thick, line width=1pt] (0,0.15) -- (0,-0.15);   
  \draw[->, thick, line width=1pt] (0,0.85) -- (0,1.15);    
}

\drawstate{1.6}{$|v\rangle$}{
  \draw[->, thick, line width=1pt] (-0.05,0.15) -- (-0.05,-0.15); 
  \draw[->, thick, line width=1pt] ( 0.05,-0.15) -- ( 0.05,0.15); 
}

\drawstate{3.2}{$|m\rangle$}{
  \draw[->, thick, line width=1pt] (0,0.15) -- (0,-0.15);   
  \draw[->, thick, line width=1pt] (1,-0.15) -- (1,0.15);   
}

\drawstate{4.8}{}{%
  \draw[->, thick] (0,0) -- (0,1);
  \draw[->, thick] (0,1) -- (1,0);
  \draw[->, thick] (1,0) -- (0,0);
}
\drawstate{6.4}{}{
  \draw[->, thick] (0,0) -- (1,0);
  \draw[->, thick] (1,0) -- (0,1);
  \draw[->, thick] (0,1) -- (0,0);
}
\end{tikzpicture}
    \caption{Generation of the lowest-order interference operators.
The first three panels illustrate representative two-step hopping
processes connecting the many-body states $|n\rangle$,
$|\nu\rangle$, and $|m\rangle$ generated by the exact
current--Hamiltonian commutator.
After projection onto the low-energy manifold of the Hubbard model,
these interference operators become the shortest virtual
flux-enclosing loops (right panels). The two loop orientations
acquire opposite Peierls phases, so their interference is governed
by the magnetic flux enclosed by the triangular plaquette. For finite next-nearest-neighbor hopping \(t'\), these virtual processes generate triangular flux-enclosing contributions to the Hall response at order \(t^2t'/U\), in addition to contributions from projected hopping within the low-energy manifold.}
    \label{fig:triangle}
\end{figure}

In the following, we illustrate the virtual doublon-mediated contributions generated by the Schrieffer–Wolff transformation, while recognizing that projected hopping processes can contribute to the Hall response already at zeroth order in \(1/U\).

For example, in Fig. \ref{fig:triangle} we represent three states $\ket{n}, \ket{\nu}, \ket{m}$ that can be connected by two hops with $t$ and one hop with $t'$. Keeping in mind that the hoppings carry Aharonov-Bohm phases, starting from the state $|m\rangle$, the current operator connects state $|\nu\rangle$ with a matrix element $iete^{i\phi_1}/\hbar$ in the negative $x$ direction, while the state $|\nu\rangle$ connects to $|n\rangle$ via $\hat H$ with a matrix element $iete^{i\phi_2}$ in the positive $y$ direction, giving a second order matrix element $ \frac{iet^2}{\hbar U}e^{i(\phi_1+\phi_2)}$. Then, state $|n\rangle$ connects back to state $|m\rangle$ via the $\hat j_y$ operator with a matrix element $\frac{iet'}{\hbar}e^{i\phi_3}$. The total matrix element for this process (the fourth figure in Fig. \ref{fig:triangle}) is
$$
-\frac{e^2t^2t'}{\hbar^2 U}e^{i(\phi_1+\phi_2+\phi_3)}.
$$
The accumulated phase $\phi_1+\phi_2+\phi_3$ is equivalent to half the unit flux $\Phi=eBa^2/2\hbar$, up to an orientation sign, through the square plaquette with area $a^2$. Subtracting the time-reversed path (fifth figure in Fig. \ref{fig:triangle}) gives the overall matrix element
\[
\textrm{Im}\!\left[\hat j^{x,\text{eff}}_{n m} \hat j^{y,\text{eff}}_{m n}\right]
=
-\frac{e^2t^2t'}{\hbar^2 U}C_\triangle\sin(\Phi/2),
\]
where $C_\triangle$ is the number of triangular loops of a fixed orientation (e.g. clockwise) per site. For a square lattice, $C_\triangle=2$, reflecting the two inequivalent triangular loops of a given orientation emanating from each lattice site when both diagonal hoppings $t'$ are present.

\begin{figure}
    \centering
\begin{tikzpicture}[scale=0.8, >=stealth]

\newcommand{\drawstate}[3]{
  \begin{scope}[shift={(#1,0)}]
    \draw[dotted] (0,0) -- (1,0) -- (1,1) -- (0,1) -- cycle;
    #3
    \node at (0.5,-0.4) {#2};
  \end{scope}
}

\drawstate{0}{$|\nu\rangle$}{
  \draw[->, thick, line width=1pt] (-0.05,0.15) -- (-0.05,-0.15);
  \draw[->, thick, line width=1pt] ( 0.05,-0.15) -- ( 0.05,0.15);
}

\drawstate{1.6}{$|n\rangle$}{
  \draw[->, thick, line width=1pt] (0,0.15) -- (0,-0.15);
  \draw[->, thick, line width=1pt] (0,0.85) -- (0,1.15);
}

\drawstate{3.2}{$|\nu'\rangle$}{
  \draw[->, thick, line width=1pt] (0,0.15) -- (0,-0.15);
  \draw[->, thick, line width=1pt] (1,0.85) -- (1,1.15);
}

\drawstate{4.8}{$|m\rangle$}{
  \draw[->, thick, line width=1pt] (0,0.15) -- (0,-0.15);
  \draw[->, thick, line width=1pt] (1,-0.15) -- (1,0.15);
}

\drawstate{6.6}{}{
  \draw[->, thick] (0,0) -- (0,1);
  \draw[->, thick] (0,1) -- (1,1);
  \draw[->, thick] (1,1) -- (1,0);
  \draw[->, thick] (1,0) -- (0,0);
}

\drawstate{8.2}{}{
  \draw[->, thick] (0,0) -- (1,0);
  \draw[->, thick] (1,0) -- (1,1);
  \draw[->, thick] (1,1) -- (0,1);
  \draw[->, thick] (0,1) -- (0,0);
}

\end{tikzpicture}
    \caption{Representative higher-order interference operators obtained after
additional Liouvillian evolution.
The left panels illustrate four-step virtual hopping processes
connecting the many-body states
$|\nu\rangle$, $|n\rangle$, $|\nu'\rangle$, and $|m\rangle$.
Following the strong-coupling projection, these operators reduce to
extended flux-enclosing loops such as square and elongated plaquettes
(right panels). These longer virtual processes provide additional flux-sensitive contributions to the Hall response. Their appearance in particular inverse-temperature coefficients depends on the full Liouvillian matrix elements and cannot be inferred from loop length alone.}
    \label{fig:square}
\end{figure}
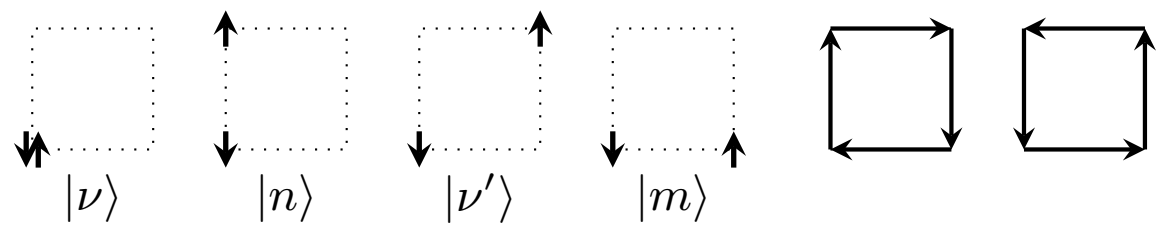

There are several combinations that contribute with four hops from the second term in Eq. (\ref{eq:v_pert}). For $t'=0$, these terms are depicted in Figure \ref{fig:square}. These processes enclose a square of area $a^2$, carrying twice the flux $2\Phi$ as the triangular plaquette. The matrix elements yield
\begin{align*}    
\textrm{Im}\!\left[\hat j^{x,\text{eff}}_{n m} \hat j^{y,\text{eff}}_{m n}\right]
=-\frac{e^2t^4}{\hbar^2 U^2} C_\square\sin(\Phi)
\end{align*}
where $C_\square$ is the number of same-oriented square plaquettes per site. For a square lattice, $C_\square=1$.

The triangular and square loops discussed above represent only the lowest-order members of the projected interference hierarchy. Additional virtual processes involving larger loops and mixed $t$--$t'$ hopping contribute at higher order, with amplitudes proportional to $t^2t'^2/U^2$, $t'^4/U^2$, and beyond. Representative examples of these higher-order interference processes are collected in Appendix~\ref{appendixb}. 

\section{Emergent coherence and the hierarchy of Hall operators}

The operator expansion derived in Section II is
formally exact and expresses the Hall conductivity entirely in
terms of equilibrium expectation values of Liouvillian
operators. 
Here, we draw attention to the important physics stemming from the virtual many-body excitations of the high temperature expansion coefficients $C_n$. The matrix elements identified above do not contribute equally to $C_n$, and are instead weighted by inverse excitation energies via the inverse Liouvillian appearing in Eqs. (\ref{eq:C1})-(\ref{eq:C3}). 

To understand the spectral weighting of the interference processes, we examine the
action of the inverse Liouvillian in the eigenbasis of the Hamiltonian. The $k=-1$ term in Eq. \eqref{eq:jj_matrix} is given by
\begin{equation}
\langle m|\mathcal{L}^{-1}A|n\rangle
=
\frac{\langle m|A|n\rangle}
{E_m-E_n},
\label{eq:Linverse}
\end{equation}
whenever \(E_m\neq E_n\). 

Thus, the inverse Liouvillian weights the Hall response by the inverse energy differences between many-body eigenstates connected by the current operators. These energy denominators encode the excitation spectrum of the correlated system, while the corresponding matrix elements determine the relative contributions of different flux-sensitive interference processes. In particular, low-energy excitations can make enhanced contributions to the Hall response, subject to cancellations between matrix elements of different signs.

This spectral sensitivity provides a possible connection to the coherence crossover observed in previous DQMC calculations~\cite{zhang_hall_2026-1}. There, the onset of integer power-law behavior coincides with a characteristic temperature extracted independently from the evolution of double occupancy. The present formulation suggests that the Hall response is sensitive to the redistribution of spectral weight as coherent low-energy electronic dynamics develop.

An important distinction arises between the formal validity of the inverse-temperature expansion and the temperature regime in which integer power laws are observed numerically. The expansion derived in Sec. \ref{sec:Louie} is controlled at sufficiently high temperatures, whereas the Schrieffer–Wolff transformation identifies the underlying virtual hopping processes independently of this thermal expansion, provided the interaction strength is sufficiently large. Remarkably, the integer power laws observed in DQMC emerge upon cooling below a characteristic coherence temperature, outside the regime in which a low-order inverse-temperature expansion is necessarily controlled.

This observation suggests that the interference processes identified by the strong-coupling analysis remain relevant as coherent low-energy dynamics develop. In particular, the suppression of certain flux-enclosing processes by lattice geometry or symmetry may favor different contributions to the Hall response over distinct temperature intervals. The correspondence between these processes and the observed integer exponents should therefore be regarded as a proposed microscopic interpretation, rather than a direct consequence of the high-temperature expansion. Establishing this correspondence quantitatively requires determining how the associated dynamical correlations evolve below the hopping scale.

The strong-coupling expansion of Sec. \ref{sec:largeU} identifies the characteristic amplitudes of the lowest-order flux-enclosing interference processes. Triangular loops involving nearest- and next-nearest-neighbor hopping have amplitudes proportional to $t^2t'/U$, while square loops generated by virtual hopping processes have amplitudes proportional to $t^4/U^2$. These amplitudes characterize the microscopic processes contributing to the Hall response, but do not uniquely determine the inverse-temperature coefficients $C_n$. The latter depend additionally on the many-body excitation spectrum and the corresponding current matrix elements. Consequently, different classes of interference processes may contribute to multiple coefficients in the inverse-temperature expansion, and their relative importance cannot be inferred from the hopping amplitudes alone.

\section{Discussion}
\label{sec:discussion}
In weakly interacting metals, the Hall effect is commonly interpreted as a measure of carrier density, Fermi-surface curvature, or scattering anisotropy. This interpretation can become incomplete in strongly correlated regimes where a quasiparticle description is no longer adequate. The Hall conductivity becomes a hierarchy of thermal expectation values of oriented interference operators. The resulting formulation identifies flux-sensitive many-body interference processes that contribute to the temperature dependence of the Hall response, without requiring a phenomenological description in terms of transport lifetimes. 

Starting from the Kubo formalism, the Liouvillian representation organizes the response in terms of current operators and their evolution under the interacting Hamiltonian. In the strong-coupling regime, the Schrieffer–Wolff transformation provides a microscopic interpretation of these processes in terms of projected hopping trajectories enclosing magnetic flux. Their contributions depend on lattice geometry, interaction strength, filling, and the correlated many-body excitation spectrum.

The operator hierarchy derived above clarifies why longitudinal and transverse transport exhibit qualitatively different temperature dependences in strongly correlated metals. Although both responses originate from the same Kubo formalism, they probe fundamentally different classes of many-body correlation functions. The longitudinal conductivity measures the ability of charge to propagate through the lattice. A single hopping event contributes directly to charge transport, so longitudinal conductivity remains finite whenever mobile carriers exist. On the other hand, the transverse response is additionally sensitive to the phases accumulated along electronic hopping trajectories. Because the magnetic field couples only through gauge-invariant phase differences, an isolated hopping process cannot contribute. Instead, Hall transport requires interference between alternative hopping trajectories enclosing finite magnetic flux. This mechanism provides a microscopic basis for the enhanced sensitivity of the Hall response to many-body coherence. While longitudinal transport can remain substantial in regimes with strongly damped electronic motion, the Hall response depends additionally on flux-sensitive correlations between hopping trajectories. Changes in these correlations can produce pronounced variations in the Hall conductivity without corresponding changes in the longitudinal response.

Our results provide a possible microscopic interpretation of the integer power laws observed in DQMC, while highlighting an important open problem: why the operator structures identified by an asymptotic inverse-temperature expansion appear to remain relevant at substantially lower temperatures. Resolving this connection may provide a route toward understanding the emergence of coherent Hall transport in correlated metals without relying on quasiparticle-based descriptions.

\begin{acknowledgments}
This work was supported by the U.S. Department of Energy (DOE), Office of Basic Energy Sciences,
Division of Materials Sciences and Engineering. E.Z.Z. acknowledges support from the Geballe Laboratory of Advanced Materials Postdoctoral Fellowship. 
\end{acknowledgments}

\appendix
\begin{widetext}
\section{Details of the Liouvillian Representation of the Hall Conductivity}
\label{appendixc}

The Hall conductivity is related to the retarded current-current correlator by
\begin{align}
\sigma_{xy}(z)
&=
\frac{i}{z}\chi_{xy}^R(z),
\end{align}
and for the Hall conductivity we consider the antisymmetric combination
\begin{equation}
\sigma^H_{xy}(z)
=
\frac{1}{2}\left[\sigma_{xy}(z)-\sigma_{yx}(z)\right].
\end{equation}
Using the Liouvillian formalism introduced in the main text, the full expression for the Hall conductivity is given by
\begin{equation}
\sigma^H_{xy}(z)
=
\frac{i}{2z}
\left\{
\left\langle
\hat j_x \frac{1}{z-\hat{\mathcal L}}
\left(1-e^{-\beta \hat{\mathcal L}}\right)
\hat j_y
\right\rangle_\beta
-
\left\langle
\hat j_y \frac{1}{z-\hat{\mathcal L}}
\left(1-e^{-\beta \hat{\mathcal L}}\right)
\hat j_x
\right\rangle_\beta
\right\}.
\end{equation}
In the energy eigenbasis,
$
({\cal \hat L}\hat j_y)_{mn}
=
(E_m-E_n)(\hat j_y)_{mn},
$
so that
$
\frac{1-e^{-\beta{\cal \hat L}}}{{\cal \hat L}}
$
is the operator analogue of the Lehmann thermal factor
$
\frac{e^{-\beta E_n}-e^{-\beta E_m}}
{E_m-E_n}.$ Inserting a complete set of eigenstates, \(H|n\rangle=E_n|n\rangle\), and defining
\[
\omega_{mn}=E_m-E_n,
\]
gives
\begin{equation}
\sigma^H_{xy}(z)
=
-\frac{1}{zZ}
\sum_{n,m}
\frac{e^{-\beta E_n}-e^{-\beta E_m}}
     {z-\omega_{mn}}
\,
\mathrm{Im}~\!\left[
j^x_{nm}j^y_{mn}
\right].
\end{equation}
The apparent singularity at \(z=0\) is removed by pairing each term
\((n,m)\) with its partner \((m,n)\). Since $
\omega_{nm}=-\omega_{mn},~~~
\mathrm{Im}\!\left[j^x_{mn}j^y_{nm}\right]
=
-\mathrm{Im}\!\left[j^x_{nm}j^y_{mn}\right]$,
we may write
\begin{equation}
\sigma^H_{xy}(z)
=
-\frac{1}{2zZ}
\sum_{n,m}
\left(e^{-\beta E_n}-e^{-\beta E_m}\right)
\mathrm{Im}~\!\left[
j^x_{nm}j^y_{mn}
\right]
\left[
\frac{1}{z-\omega_{mn}}
+
\frac{1}{z+\omega_{mn}}
\right].
\end{equation}
The expression in brackets contains an explicit factor of \(z\):
\begin{equation}
\frac{1}{z-\omega}
+
\frac{1}{z+\omega}
=
\frac{2z}{z^2-\omega^2}.
\end{equation}
Therefore the external \(1/z\) cancels, and the dc limit,
$
\sigma_{xy}^H
=\lim_{z\rightarrow 0}
\frac{1}{2}
\left[
\sigma_{xy}(z)-\sigma_{yx}(z)
\right]$
is finite:
\begin{align}
\sigma^H_{xy}
=
\frac{1}{Z}
\sum_{n,m}
\frac{
e^{-\beta E_n}-e^{-\beta E_m}
}
{\omega_{mn}^2}
\,
\mathrm{Im}~\!\left[
j^x_{nm}j^y_{mn}
\right].
\end{align}
Expanding the Boltzmann difference,
\begin{align}
e^{-\beta E_n}-e^{-\beta E_m}
=
e^{-\beta E_n}
\left(1-e^{-\beta\omega_{mn}}\right)
=
e^{-\beta E_n}
\sum_{p=1}^{\infty}
\frac{(-1)^{p+1}\beta^p}{p!}
\omega_{mn}^{p},
\end{align}
we recover the dc high-temperature expansion
\begin{align}
\sigma^H_{xy}
=
\sum_{p=1}^{\infty}
\frac{(-1)^{p+1}\beta^p}{p!}
\,
\mathrm{Im}~
\left\langle
\hat j_x
\hat{\mathcal L}^{p-2}
\hat j_y
\right\rangle_\beta .
\end{align}
The first few terms are therefore
\begin{align}
\sigma^H_{xy}
&=
\beta\,
\mathrm{Im}
\left\langle
\hat j_x
\hat{\mathcal L}^{-1}
\hat j_y
\right\rangle_\beta
-\frac{\beta^2}{2}
\mathrm{Im}
\left\langle
\hat j_x
\hat j_y
\right\rangle_\beta
+\frac{\beta^3}{6}
\mathrm{Im}
\left\langle
\hat j_x
\hat{\mathcal L}
\hat j_y
\right\rangle_\beta
+O(\beta^4)
\\
&=
\beta\,
\mathrm{Im}
\left\langle
\hat j_x
\hat{\mathcal L}^{-1}
\hat j_y
\right\rangle_\beta
-\frac{\beta^2}{2}
\mathrm{Im}
\left\langle
\hat j_x
\hat j_y
\right\rangle_\beta
+\frac{\beta^3}{6}
\mathrm{Im}
\left\langle
\hat j_x
[\hat H,\hat j_y]
\right\rangle_\beta
+O(\beta^4)\\
&=
\beta\,
\mathrm{Im}
\left\langle
\hat j_x
\hat{\mathcal L}^{-1}
\hat j_y
\right\rangle_\beta
-\frac{\beta^2}{4i}
\left\langle
[\hat j_x,\hat j_y]
\right\rangle_\beta
+\frac{\beta^3}{12i}
\left\langle
[\hat j_x,[\hat H,\hat j_y]]
\right\rangle_\beta
+O(\beta^4).
\end{align}

This last form makes the Hall nature explicit: every term is written as the expectation value of an antisymmetric commutator structure. One obtains schematically
\begin{align}
\sigma_{xy}^H
=
\beta C_1
+
\beta^2 C_2
+
\beta^3 C_3
+\cdots ,
\end{align}
where
\begin{align}
C_1
&=
\mathrm{Im}
\left\langle\hat
j_x
\mathcal{\hat L}^{-1} \hat j_y
\right\rangle_\beta,\\
C_2
&=-\frac{1}{2}\mathrm{Im}
\left\langle\hat
j_x\hat j_y
\right\rangle_\beta,\\
C_3
&=\frac{1}{6}\mathrm{Im}
\left\langle\hat
j_x\mathcal{\hat L} \hat j_y
\right\rangle_\beta.
\end{align}

\section{Higher-Order Virtual Interference Processes}
\label{appendixb}
The discussion in Sec.~\ref{sec:largeU} concentrated on the shortest virtual
interference processes because these determine the leading
contributions to the Hall response and most clearly illustrate the
physical origin of the operator hierarchy. The complete
strong-coupling expansion, however, generates many additional
interference operators involving larger virtual loops and mixed
nearest- and next-nearest-neighbour hopping processes.

Representative examples of these higher-order virtual interference
paths are collected here. Although their amplitudes are suppressed
by additional powers of $t/U$ or $t'/U$, they provide the
microscopic origin of the higher-order coefficients in the
inverse-temperature expansion and demonstrate that the Hall
operator naturally organizes itself into an increasingly rich
hierarchy of flux-enclosing many-body processes.

The examples collected in this Appendix are not intended as an
exhaustive catalog of virtual hopping processes. Instead, they
illustrate the systematic manner in which successive Liouvillian
evolution generates increasingly extended interference operators.
The emergence of this hierarchy is independent of the particular
lattice geometry and reflects the general organization of Hall
transport in terms of coherent many-body interference.

For $t'$ only, a diamond path is created with area $2a^2$, depicted in Figure \ref{fig:diamond}.
The matrix elements yield
\begin{align}
\textrm{Im}\!\left[\hat j^{x,\text{eff}}_{n m} \hat j^{y,\text{eff}}_{m n}\right]=-\frac{e^2t'^4}{\hbar^2 U^2} C_\diamond\sin(2\Phi),
\end{align}
where $C_\diamond$ is the number of same-oriented diamond plaquettes per site. For a square lattice, $C_\diamond=2$.

Combinations of $t$ and $t'$ creates either flattened triangles or trapezoids, each enclosing an area $a^2$. The trapezoids are depicted in Figure \ref{fig:traps}. The matrix elements yield
\begin{align}
   \textrm{Im}\!\left[\hat j^{x,\text{eff}}_{n m} \hat j^{y,\text{eff}}_{m n}\right]=-
\frac{e^2t^2t'^2}{\hbar^2 U^2} C_{\text{trap}}\sin(\Phi),
\end{align}
where $C_{\text{trap}}$ is the number of same-oriented trapezoidal plaquettes per site. For a square lattice, $C_{\text{trap}}=4$.

The flattened triangles are depicted in Figure \ref{fig:flat}. The matrix elements yield
\begin{align}
    \textrm{Im}\!\left[j^{x,\text{eff}}_{n m} j^{y,\text{eff}}_{m n}\right] =- \frac{e^2t^2t'^2}{\hbar^2 U^2} C_{\text{flat}}\sin(\Phi),
\end{align}
where $C_{\text{flat}}$ is the number of same-oriented flattened triangle loops per site. For a square lattice, $C_{\text{flat}}=4$.

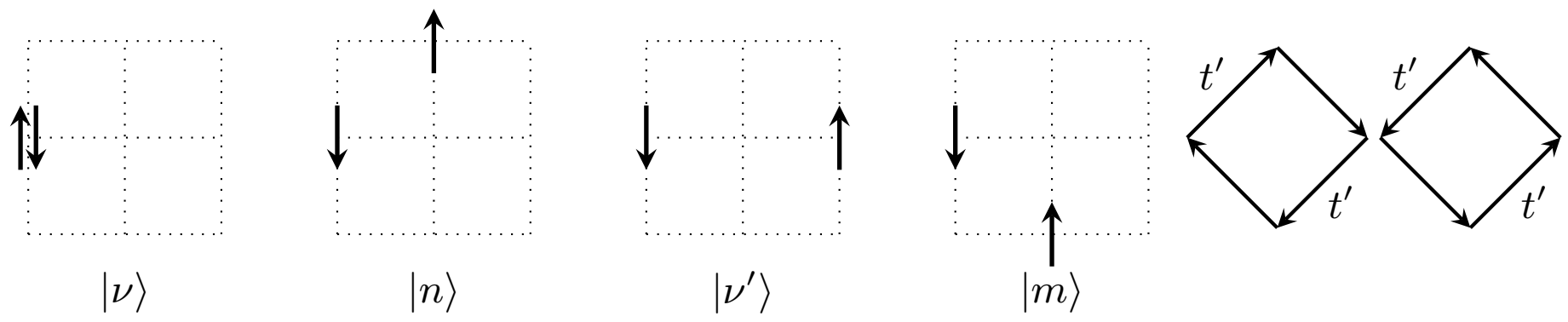
\begin{figure}
    \centering
\begin{tikzpicture}[scale=1.0, >=stealth]
\newcommand{\drawstate}[3]{
  \begin{scope}[shift={(#1,0)}]

    \foreach \x in {0,0.75,1.5}
      \draw[dotted] (\x,0) -- (\x,1.5);
    \foreach \y in {0,0.75,1.5}
      \draw[dotted] (0,\y) -- (1.5,\y);

    #3

    \node at (0.75,-0.45) {#2};

  \end{scope}
}

\newcommand{\drawdiamondcw}[1]{
  \begin{scope}[shift={(#1,0.75)}]
    \draw[->, thick] (0,0) -- (0.7,0.7);
    \draw[->, thick] (0.7,0.7) -- (1.4,0);
    \draw[->, thick] (1.4,0) -- (0.7,-0.7);
    \draw[->, thick] (0.7,-0.7) -- (0,0);
    \node at (0.2,0.5) {$t'$};
    \node at (1.2,-0.5) {$t'$};
  \end{scope}
}

\newcommand{\drawdiamondccw}[1]{
  \begin{scope}[shift={(#1,0.75)}]
    \draw[->, thick] (0,0) -- (0.7,-0.7);
    \draw[->, thick] (0.7,-0.7) -- (1.4,0);
    \draw[->, thick] (1.4,0) -- (0.7,0.7);
    \draw[->, thick] (0.7,0.7) -- (0,0);
    \node at (0.2,0.5) {$t'$};
    \node at (1.2,-0.5) {$t'$};
  \end{scope}
}

\drawstate{0}{$|\nu\rangle$}{
  \draw[->, thick, line width=1pt] (-0.06,0.5) -- (-0.06,1.0);
  \draw[->, thick, line width=1pt] ( 0.06,1.0) -- ( 0.06,0.5);
}%

\drawstate{2.4}{$|n\rangle$}{
\draw[->, thick, line width=1pt] (0.75,1.25) -- (0.75,1.75);
\draw[->, thick, line width=1pt] ( 0.0,1.0) -- ( 0.0,0.5);
}%

\drawstate{4.8}{$|\nu'\rangle$}{
  \draw[->, thick, line width=1pt] (1.5,0.5) -- (1.5,1.0);
  \draw[->, thick, line width=1pt] ( 0.0,1.0) -- ( 0.0,0.5);
}%

\drawstate{7.2}{$|m\rangle$}{
  \draw[->, thick, line width=1pt] (0.75,-0.25) -- (0.75,0.25);
  \draw[->, thick, line width=1pt] ( 0.0,1.0) -- ( 0.0,0.5);
}%

\drawdiamondcw{9}
\drawdiamondccw{10.5}
\end{tikzpicture}
\caption{Representative four-step interference processes generated by the
operator hierarchy in the presence of next-nearest-neighbour hopping
$t'$. The left panels illustrate the sequence of virtual intermediate
many-body states $|\nu\rangle$, $|n\rangle$, $|\nu'\rangle$, and
$|m\rangle$. After projection onto the low-energy manifold these
processes become oriented diamond-shaped virtual loops (right),
providing an additional class of flux-sensitive interference operators
that contributes to the higher-order Hall coefficients.}
    \label{fig:diamond}
\end{figure}

\begin{figure}
    \centering
\begin{tikzpicture}[scale=1.0, >=stealth]

\newcommand{\drawstate}[3]{
  \begin{scope}[shift={(#1,0)}]

    \foreach \x in {0,0.75,1.5}
      \draw[dotted] (\x,0) -- (\x,1.5);
    \foreach \y in {0,0.75,1.5}
      \draw[dotted] (0,\y) -- (1.5,\y);

    #3

    \node at (0.75,-0.45) {#2};

  \end{scope}
}

\newcommand{\drawtrapcw}[1]{
  \begin{scope}[shift={(#1,-0.25)}] 
    \draw[->, thick] (0,0.75) -- (0.75,1.5);
    \draw[->, thick] (0.75,1.5) -- (1.5,1.5);
    \draw[->, thick] (1.5,1.5) -- (0.75,0.75);
    \draw[->, thick] (0.75,0.75) -- (0,0.75);
  \end{scope}
}

\newcommand{\drawtrapccw}[1]{
  \begin{scope}[shift={(#1,-0.25)}]
    \draw[->, thick] (0.75,1.5) -- (0,0.75);
    \draw[->, thick] (1.5,1.5) -- (0.75,1.5);
    \draw[->, thick] (0.75,0.75) -- (1.5,1.5);
    \draw[->, thick] (0,0.75) -- (0.75,0.75);
  \end{scope}
}

\drawstate{0}{$|\nu\rangle$}{
  \draw[->, thick, line width=1pt] (-0.06,0.5) -- (-0.06,1.0);
  \draw[->, thick, line width=1pt] ( 0.06,1.0) -- ( 0.06,0.5);
}%

\drawstate{2.0}{$|n\rangle$}{
\draw[->, thick, line width=1pt] (0.75,1.25) -- (0.75,1.75);
\draw[->, thick, line width=1pt] ( 0.0,1.0) -- ( 0.0,0.5);
}%

\drawstate{4.4}{$|v'\rangle$}{
  \draw[->, thick, line width=1pt] (1.5,1.25) -- (1.5,1.75);
\draw[->, thick, line width=1pt] ( 0.0,1.0) -- ( 0.0,0.5);
}%

\drawstate{6.8}{$|m\rangle$}{
  \draw[->, thick, line width=1pt] (0.75,0.5) -- (0.75,1.0);
  \draw[->, thick, line width=1pt] ( 0.0,1.0) -- ( 0.0,0.5);
}%

  \drawtrapcw{9}
  \drawtrapccw{10.5}

\end{tikzpicture}
\caption{Another family of higher-order interference operators generated by
successive Liouvillian evolution. The corresponding strong-coupling
representation is an elongated four-site virtual loop (right). These
extended interference paths appear naturally as higher members of the
operator hierarchy and contribute to the coefficients beyond the
leading Hall term.}
    \label{fig:traps}
\end{figure}
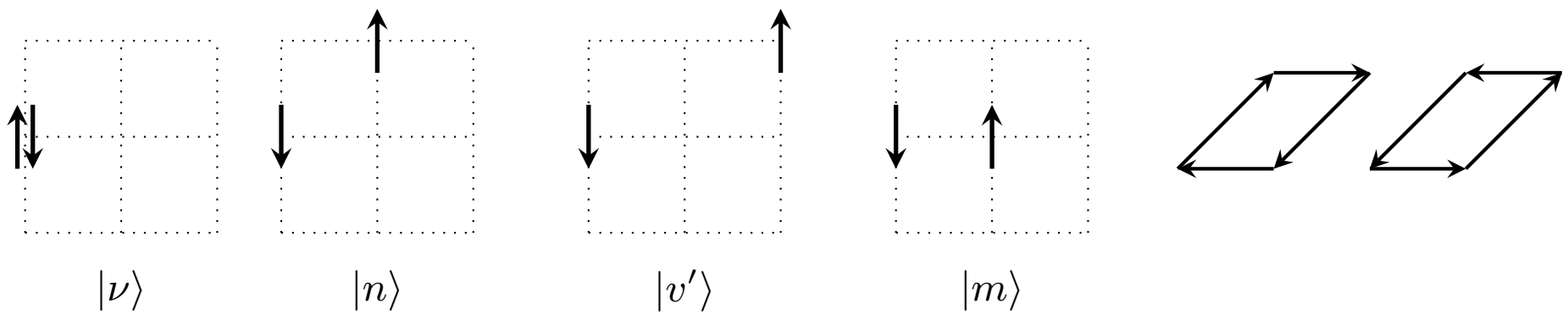

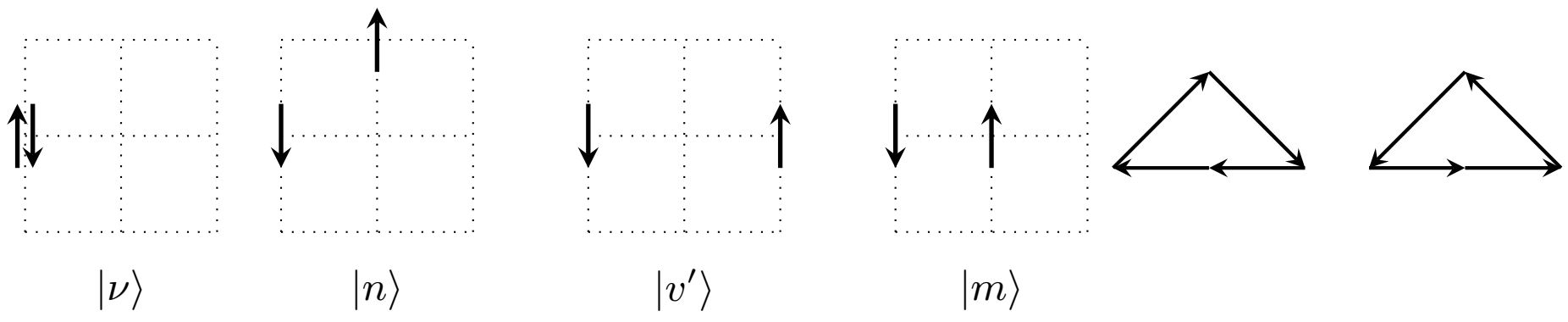
\begin{figure}
    \centering
\begin{tikzpicture}[scale=1.0, >=stealth]

\newcommand{\drawstate}[3]{
  \begin{scope}[shift={(#1,0)}]

    \foreach \x in {0,0.75,1.5}
      \draw[dotted] (\x,0) -- (\x,1.5);
    \foreach \y in {0,0.75,1.5}
      \draw[dotted] (0,\y) -- (1.5,\y);

    #3

    \node at (0.75,-0.45) {#2};

  \end{scope}
}

\newcommand{\drawflatcw}[1]{
  \begin{scope}[shift={(#1,-0.25)}] 
    \draw[->, thick] (0,0.75) -- (0.75,1.5);
    \draw[->, thick] (0.75,1.5) -- (1.5,0.75);
    \draw[->, thick] (1.5,0.75) -- (0.75,0.75);
    \draw[->, thick] (0.75,0.75) -- (0,0.75);
  \end{scope}
}

\newcommand{\drawflatccw}[1]{
  \begin{scope}[shift={(#1,-0.25)}]
    \draw[->, thick] (0.75,1.5) -- (0,0.75);
    \draw[->, thick] (1.5,0.75) -- (0.75,1.5);
    \draw[->, thick] (0.75,0.75) -- (1.5,0.75);
    \draw[->, thick] (0,0.75) -- (0.75,0.75);
  \end{scope}
}

\drawstate{0}{$|\nu\rangle$}{
  \draw[->, thick, line width=1pt] (-0.06,0.5) -- (-0.06,1.0);
  \draw[->, thick, line width=1pt] ( 0.06,1.0) -- ( 0.06,0.5);
}%

\drawstate{2.0}{$|n\rangle$}{
\draw[->, thick, line width=1pt] (0.75,1.25) -- (0.75,1.75);
\draw[->, thick, line width=1pt] ( 0.0,1.0) -- ( 0.0,0.5);
}%

\drawstate{4.4}{$|v'\rangle$}{
  \draw[->, thick, line width=1pt] (1.5,0.5) -- (1.5,1.0);
\draw[->, thick, line width=1pt] ( 0.0,1.0) -- ( 0.0,0.5);
}%

\drawstate{6.8}{$|m\rangle$}{
  \draw[->, thick, line width=1pt] (0.75,0.5) -- (0.75,1.0);
  \draw[->, thick, line width=1pt] ( 0.0,1.0) -- ( 0.0,0.5);
}%

  \drawflatcw{8.5}
  \drawflatccw{10.5}

\end{tikzpicture}
    \caption{Lowest-order flux-sensitive interference operator generated by the
strong-coupling projection of the exact Hall operator in the presence
of next-nearest-neighbour hopping $t'$. The left panels illustrate the
sequence of virtual intermediate many-body states connecting the
initial and final configurations. After projection into the
low-energy manifold, the process reduces to an oriented triangular
virtual hopping loop (right). The two opposite loop orientations
acquire opposite Peierls phases and therefore contribute with opposite
signs to the Hall conductivity. This is the leading interference
process responsible for the coefficient $C_1$, with an amplitude of
order $t^2t'/U$.}
    \label{fig:flat}
\end{figure}

Finally, we can sum all of the contributions into one expression:
\begin{align}
    \textrm{Im}\!\left[j^{x,\text{eff}}_{n m} j^{y,\text{eff}}_{m n}\right]=-\frac{e^2}{\hbar^2 U} \Biggl[&C_\triangle~t't^2\sin(\Phi/2)+    
    C_\square \frac{t^4}{U}\sin(\Phi)+C_\diamond \frac{t'^4}{U}\sin(2\Phi)+(C_{\text{trap}}+C_{\text{flat}})\frac{t^2t'^2}{U}\sin(\Phi)\Biggr].
\end{align}
Since all coefficients $C$ are of order 1, the triangular term dominates and the other terms are smaller by factors of $t/U, ~t'/U$.
\end{widetext}

We can summarize these coefficients in Table \ref{tab:table}. This approach extends to other lattices (honeycomb, kagome, etc.) or other models (extended Hubbard, etc.). 

\begin{table}[]
    \centering
    \begin{tabular}{|c|c|c|c|c|c|}\hline
Loop Type &	Count &	Hops &	Flux &	Scaling &	Sign(t') dep? \\\hline\hline
3-hop triangles &	$C_\triangle = 2$	& $t^2t'$&	$\Phi/2$ &	$1/U$ &	Yes\\\hline
Square &	$C_\square = 1$ &	$t^4$ &	$\Phi$ &	$1/U^2$	& No \\\hline
Diamond &	$C_\diamond = 2$	&$t'^4$ &	$2\Phi$ &	$1/U^2$	& No \\\hline
Trapezoid &	$C_{\text{trap}} = 4$ &	$t^2t'^2$ &	$\Phi$ &	$1/U^2$	&No \\\hline
Flat triangle &	$C_{\text{flat}} = 4$ &	$t^2t'^2$ &	$\Phi$ &	$1/U^2$ &	No\\\hline
    \end{tabular}
    \caption{Characterization of the Hall response matrix elements in the strongly correlated regime $t\ll U$.}
    \label{tab:table}
\end{table}

\section{Interference Structure of the Kubo Formula}
\label{appendixd}

The main text reformulates the Hall conductivity in terms of a Liouvillian hierarchy of interference operators.  It is important to emphasize, however, that this hierarchy does not introduce new physics beyond linear-response theory.  Rather, it reorganizes the Kubo formula in a manner that makes the microscopic origin of Hall transport transparent.  In this Appendix we briefly show how the interference picture is already contained in the Kubo expression and how the infinite-temperature limit naturally leads to the operator hierarchy developed in the main text.

The real part of the dc Hall conductivity may be written in Lehmann representation as
\begin{equation}
\mathrm{Re}\,
\sigma_{\alpha\beta}
=
\frac{\hbar}{VZ}
\sum_{nm}
\frac{
e^{-\beta E_n}
-
e^{-\beta E_m}
}
{(E_n-E_m)^2}
\,
\mathrm{Im}
\!\left[
\langle n|J_\alpha|m\rangle
\langle m|J_\beta|n\rangle
\right].
\label{eq:lehmann_appendix}
\end{equation}

Defining
\begin{equation*}
\Delta_{mn}=E_m-E_n,
\end{equation*}
we expand the the thermal factor in Eq.~(\ref{eq:lehmann_appendix}) by
\begin{equation}
\frac{
e^{-\beta E_n}-e^{-\beta E_m}
}{
(E_m-E_n)^2
}
=
\frac{\beta}{\Delta_{mn}}
-\frac{\beta^2}{2}
+\frac{\beta^3}{6}\Delta_{mn}
+\mathcal O(\beta^4).
\label{eq:gibbs}
\end{equation}
Substituting this series into Eq.~(\ref{eq:lehmann_appendix}) recovers the high-temperature interference hierarchy in Eq.~(\ref{eq:master}). 

Unlike the longitudinal conductivity, which depends upon the modulus squared of a single current matrix element, the Hall conductivity is determined by the imaginary part of the product of two current operators.  The Hall response therefore compares two successive current-induced hopping amplitudes and is intrinsically sensitive to their relative quantum phase.  In a magnetic field, this phase is gauge invariant and is equal to the magnetic flux enclosed by the competing hopping trajectories.  The Hall effect is therefore an interference phenomenon already at the level of the Kubo formula.

The Liouvillian formulation introduced in the main text reorganizes Eq.~(\ref{eq:lehmann_appendix}) by expressing the response in terms of repeated commutators of the current operator with the Hamiltonian. The dynamics are encoded in the hierarchy
\begin{equation}
J_y,\qquad
{\cal L}J_y,\qquad
{\cal L}^2J_y,\qquad
{\cal L}^3J_y,\ldots ,
\end{equation}
which systematically generates interference operators of increasing spatial complexity. 

Only operator products whose hopping sequences return every particle to its original configuration possess a nonzero trace.  Open hopping strings therefore vanish identically, whereas closed oriented hopping loops survive.  The imaginary part of Eq.~(\ref{eq:lehmann_appendix}) further selects the component that is odd under reversal of the loop orientation, making the Hall response sensitive only to flux-enclosing interference processes.

The expansion Eq. \ref{eq:gibbs} organizes the Hall conductivity into a hierarchy of operator traces involving progressively more complicated closed interference paths.  If the lowest-order trace vanishes because of a symmetry, such as particle-hole symmetry on a bipartite lattice, the leading Hall response is pushed to the next order in $\beta$. Consequently, the dominant temperature dependence changes from $1/T$ to $1/T^2$, $1/T^3$, and so forth, producing the integer power laws discussed in the main text.

The Liouvillian hierarchy therefore provides an exact operator reorganization of the Kubo formula. It exposes the microscopic objects measured by the Hall effect, which is a hierarchy of oriented many-body interference operators whose strong-coupling realization is given by virtual flux-enclosing hopping loops.

\bibliography{main}

\end{document}